\documentclass{article}

\usepackage[english]{babel}
\usepackage{eurosym}

\usepackage[usenames]{color}

\begin{document}

\title{Decision Taking as a Service%
}
\author{%
Jan A. Bergstra\\
{\small  Section Theory of Computer Science,
Informatics Institute,} \\
{\small Faculty of Science, University of Amsterdam,}\\
{\small Amsterdam, The Netherlands.}%
\thanks{Author's email address: {\tt j.a.bergstra@uva.nl}. In this first revision of the original paper from 2011
many minor improvements have been made.
}
%\date{}
}

\maketitle

\begin{abstract}
\noindent Decision taking can be performed as a service to other parties and it is 
amenable to outtasking 
rather than to outsourcing. Outtasking decision taking is compatible with 
selfsourcing of decision making activities 
carried out in preparation of decision taking. Decision taking as a service 
(DTaaS) is viewed as an instance of 
so-called decision casting. Preconditions for service casting are examined, 
and compliance of decision taking with these preconditions is confirmed. 
Potential advantages and disadvantages of using decision taking as a service are considered.
\end{abstract}

\tableofcontents

\section{Introduction}\label{sec:Intro}
This paper has been written with the objective to state and answer the question to what extent it is possible and meaningful to use and to offer decision taking (DT) as a service, hereafter abbreviated as DTaaS.

For the definition of decision taking that I will use, I refer to  \cite{Bergstra2011d}.  That definition is unusual in the to the extent that  a decision is an action rather than a result of an action. 

\subsection{Service casting and service casting preconditions}
DTaaS will be understood as a substitution instance of the context [-]aaS where DT is placed in the ``hole''. Thus DTaaS  abbreviates [DT]aaS. For an activity or entity or  process X, XaaS (= [X]aaS) is what I will call the service casting of X. [X]aaS is defined only if X satisfies so-called service casting requirements. 

In particular DTaaS represents the service casting of DT, provided that DT satisfies the service casting requirements, that is a collection of constraints that any X must comply with for [X]aaS to make sense, that is to be well-defined. These constraints are called service casting preconditions. The objective of this paper is twofold: (i) to determine a reasonable account of the service casting preconditions in general, and (ii) to find out why  and to what extent DT complies with these service casting preconditions.

\subsection{Parametrized roles}
S abbreviates service and [-]aaS stands for the service casting operator X $\rightarrow$ [X]aaS. S can be considered a role and the service casting operator can be understood as an instance of the so-called generic role 
casting operator which takes role Y to the role casting operator [-]aa[Y].%
\footnote{[-]aaS = (let Y = S in)[-]aa[Y].}

Below I will make use of the role casting operator [-]aa[F] where F stands for feature. XaaF views X (or instances of X) as a feature of entities or processes that incorporate some form of X. 

In principle role casting preconditions indicate for which roles Y, [-]aa[Y] is a valid role casting operator. I will avoid this level of generality and I will assume that ``service'' satisfies suitable role casting preconditions without making an attempt yo analyze these in detail.

\subsection{Decisions and units}

In \cite{Nutt2011} the observation is made that 
decision making research mainly stands on two feet: description and prescription; 
a similar dichotomy underlies the 
survey \cite{FallonButterfield2005}. 
The same may hold for work on decision taking. I will avoid both description and prescription, 
and instead focus on the structure of decision taking as behavioral processes for 
agents and for groups of agents.  
I will understand decision making and decision taking as structural mechanisms for multi-agent cooperation 
independently of existing practice. Decision taking constitutes an organizational control mechanism. 

Following \cite{BDV2011b,BDV2011c} I will speak of units to indicate all forms of organizational entities. Units may be equipped with a hierarchical structure includingg subunits at various levels. Sourcements (see \cite{BDV2011a}) are descriptions of sourcing relations between units. Outsourcing and outtasking are examples of sourcement transformations.

Decision making (DM) can be split in two parts: decision preparation (DP) and decision taking. In symbols: ``DM = DP + DT''. Because DT is usually connected with leading roles and a unit will not function without leaders, it is implausible that DT is outsourced together with the sources (leaders) responsible for it. 

Some authors (e.g. see \cite{WhiteleyWood2000}) read decision taking as the reception of a decision outcome in the role of staff responsible for realizing the goals of the decision taker. Although this is a coherent interpretation of decision taking I will deviate from it. I will not use or propose 
a special term for the action or process of taking notice and subsequent comprehension of a decision outcome. Other authors, e.g. in \cite{SheardKakabadse2007} use decision taking as an equivalent of decision making, a use of the terminology that I am not following either. 

In \cite{AtkinsonGoodman2008} decision making and decision taking are distinguished in a way that comes close to the interpretation of \cite{Bergstra2011d}  by accepting the real time character and the situational dependency of  decision taking, 
in contrast with decision making, but without assuming the convention that a decision is an action. Below I will make use of so-called outtasking of decision taking. Admittedly this is an uncommon phrase; it is related to delegation of decisions as found in \cite{SengulGD2012} and in \cite{Errington1986}. 

In \cite{Valdman2010} the outsourcing of self-governent is analyzed as a philosophical theme. The analysis leads the author to formulating doubts concerning the mentioned notion. These doubts are grounded in both the human nature of the outsourcing agent and in the comprehensiveness of the range of decisions supposed to be outsourced. In the terminology that I prefer to use, outtasking would be preferable to outsourcing for the use made of in in \cite{Valdman2010}, because the human agent, viewed as a unit, cannot possibly export any of its internal sources (given today's biotechnology at least) to an external unit. 

\subsection{Organization of the paper}
DT is contrasted with voting and promise issuing in Section \ref{vapi}, then in Section \ref{DTaaF} DT is viewed as a feature of organizational behavior. A comparison with computer program structuring is made, and it is described how an organization may proceed if it makes no use, or limited use, of the  DT feature. In Section \ref{Osasc} the concept of service casting, that is casting an entity class or an activity class (concept) as a service, is examined in rigorous detail. Several examples of service casting are considered and service casting preconditions as well as service casting obstacles are collected. 

In Section \ref{DTaaS} the particular case of service casting DT is approached by indicating how DT complies with the service casting preconditions and avoids a match with the service casting obstacles.  In Section \ref{prfuDTaaS} possible rationales for making use of DTaaS are surveyed and a preliminary risk analysis for DTaaS is provided.
In Section \ref{Cremarks} the intended audience for a theory of 
DTaaS is specified and the intended yield for those  in the intended 
audience of acquiring familiarity with our theory of DTaaS is formulated. 
Finally in Section \ref{Cremarks}  some conclusions and directions for further work are mentioned.

\section{Voting, choosing, promising, and guessing}
\label{vapi}
For the purpose of this paper the boundaries of  decision must be laid down with more precision than has been done in \cite{Bergstra2011d}. In \cite{Bergstra2011d} the distinction between decision and choice has been discussed extensively, with the effect that (i) an act of choice need not be a decision and, (ii) that a decision need not involve choice. 

When contemplating DT free activity it matters to what extent choice, voting, and promising are considered
to overlap decision.
\subsection{Formalizing aspects of ontology}
\label{ontology}
With X $\subseteq$ Y (X is a subclass of Y) I will mean that every instance of 
X is an instance of Y. X-Y denotes  the class of instances of X that are not instances of Y. When
considering classes of actions c(X), the complement of X is ACT - X.

With ACT denoting the class of actions and with CH denoting  
the class of choices (acts of choosing from a menu of alternatives) it follows that DT $\subseteq$ ACT,  
CH $\subseteq$ ACT, while DT and CH are not disjoint and neither 
one is a subclass of the other. But I maintain that most decisions are not choices.

If by default (but not necessarily always) instances of X are also instances of Y that relation 
between X and Y is denoted as X $\subseteq_d$ Y. 
It follows that X $\subseteq$ Y implies X $\subseteq_d$ Y. In particular
we find DT $\subseteq_d$ c(CH) and CH $\subseteq_d$ c(DT).

Individual voting (IV) as well as group voting (GV) are subclasses of choice, thus: IV $\subseteq$ CH, GV $\subseteq$ CH, IV $\subseteq_d$ c(DT), and GV $\subseteq_d$ c(DT).

Social choice (SC) is an aggregate of which IV's are a part (component) rather than an instance, and SC $\subseteq$ CH. Concurrent IV's are collected in a group vote (GV) which itself is embedded (used as a phase in) social choice. 
Now GV $\subseteq_d$ CH, and SC $\subseteq_d$ CH.

Promise issuing (PI)  is a  class of actions disjoint from DT, so PI  $\subseteq$ c(DT).   
Moreover PI $\subseteq_d$ c(CH), that is, most most acts of issuing a promise are not choices. 
Obligation creating promise issuings (OCPI) are a proper subclass
of PI, so OCPI $\subset$ PI.

For a class of actions X  service casting of its instances ([X]aaS) is considered plausible  if for most instances of X it is conceivable in technical terms that these are provided as a service. I will consider [X]aaS implausible if it is not the case that [X]aaS is plausible. Now I will use the (defeasible) rule that if X $\subseteq_d$ Y $\&$ pl([Y]aaS) $\Rightarrow$ pl([X]aaS). 

As an example: let DT4I denote DT for investment issues and DT4F denote DT for financial matters. 
Then DT4I $\subseteq_d$ DT4F, and if one assumes pl([DT4F]aaS) it follows that pl([DT4I]aaS) as well.

Guessing may be considered a subclass of choosing: G $\subset$ CH. Making a Random guess (RG) is a subclass of guessing. Indeed, guessing cannot constitute
a decision, but of course it may constitute the mechanism by which a choice is made on which a decision is bases. I assume that pl([G]aaS) because providing an unbiased guess is not easy and may require a form of independence that can be provided by a specialized agent to a consumer of guesses. Moreover,
RG may require specialized technology and thereofer pl([RG]aaS) can also be justified.

\subsection{Voting, social choice, and decision}

However, between choice and decision there is voting, as well as social choice. Related to decision is promising (promise issuing). I hold that voting, social choice, and promising by default (that is in most cases) do not constitute decisions. 
Clearly understanding the difference between promising, voting, social choice, and decision is necessary because neither voting, or social choice, nor promising can be cast as a service.

\subsection{Voting and social choice}
In order to distinguish voting from decision taking I will make the following assumptions about voting:

\begin{enumerate}
\item Casting a vote (by an individual voter)  is an act of voting. 
\item Each act of voting is an instance of choice.
\item The result of casting a vote (by an individual voter) is an individual voting outcome (IVO).%
\footnote{An IVO is often referred to as a vote but that terminology is confusing
as a vote may also refer to an act of voting,
which I prefer as its meaning. It may even refer to the fact that an agent has the right to vote. The common use of the term vote seems to vary between static and dynamic interpretations. This process product ambiguity is quite common and may often lead to confusion. As an example of process product ambiguity I mention the notion test in software engineering (see \cite{Middelburg2010b}).}
\item Individual voters act concurrently,%
\footnote{This form of concurrency can be adequately formalized with arbitrary interleaving based parallelism, for instance using the process algebras specified in \cite{BaetenBastenReniers2009}.}
while together performing a group voting process.
\item A group voting process terminates and subsequently the plurality of collected IVO's can be aggregated to a single AVO (aggregate voting outcome) by means of some aggregation method.%
\footnote{Making an aggregate social choice (and thus producing an AVO) as implemented by means of voting is in most cases not an instance of decision taking as understood in \cite{Bergstra2011d} because the aggregate agent has no intentions which it is aiming for and
because  it makes no predictions about the consequences of putting the AVO into effect. Social choice,  is not an instance of choice and social choice is not necessarily an instance of decision either. By default it is not. In fact, In many political systems social choice is a more powerful mechanism than decision taking.}

The design, analysis, and selection of aggregation methods for votings belongs to the area of  social choice theory.
There exists a large volume of theory on voting and social choice for instance \cite{RikerOrdeshook1968,Young1988}, most of it not in need of notions like IVO and AVO. An AVO is frequently referred to as a decision, thus reflecting a point of view that I do not share.

\item The AVO is merely the outcome of the overall voting process, and it must be distinguished from the consequences of putting it into effect.

\item Depending on the particular organization of the group of potential voters as well as depending on its objectives, casting a vote may range from being an instance of making a choice to being an instance of taking a decision, with in between acts of voting that constitute a class of their own. Here are two extremes:
\begin{itemize}
\item  If the voting outcomes cannot be traced back to individual voters (anonymity), and if the voter cannot predict the impact of his/her particular choice made when voting, the IVO will not have the form of impact the voter expects from a decision outcome, and for that reason its coming about will not count as a decision.
\item If the following conditions are satisfied:
\begin{itemize}
\item the vote is open,
\item if the voter can reasonably predict the impact of its own contribution to the aggregate outcome, 
\item the voter has a clear mind about the impact that putting the AVO into effect is likely to have, 
\item its by ``publication'' of the AVO that its putting into effect is triggered, not simply by the participants in the voting feeling permitted to go their own way,
\item the entire voting group acts on the basis of a clear role, and 
\item each member is acting in terms of that role.
\end{itemize}
then,
\begin{itemize}
\item the voting may be considered a group decision with the AVO constituting the 
corresponding decision outcome, and 
\item  the IVO may be considered a decision outcome
(in fact it is only a constituent of a decision outcome) and the preceding act of voting may be 
considered a decision taken by the individual agent.
\end{itemize}
The degree to which these considerations are consistent with the assumptions IV $\subseteq_d$ c(DT), 
and GV $\subseteq_d$ c(DT) that were formulated in Paragraph~\ref{ontology} above depends on one's attitude towards voting's in general.
\end{itemize}

\item If a group votes on which direction to proceed in a dangerous area that voting is a matter of 
choice or of action determination but not of decision taking. If a group vote is used by one or more members for the justification of their own subsequent activity than referring to the voting as an instance of decision taking goes against the idea of OODT.

\item In some cases a decision outcome needs to be confirmed (ratified) afterwards by some  body with higher authority. Then the action producing that ``decision outcome''  still counts as a decision if it was taken with the expectation that ratification will succeed. This leads to the following pipeline: 
\begin{itemize}
\item A process involving different levels and stages  of decision taking and social choice leads to the listing of 
persons that may be elected in office.  
\item Social choice mechanisms, often based on voting, are used to select who will be in office for some period. 
\item Persons in office enact decision taking, for that task they may be supported by non-elected staff for 
decision making (decision preparation). 
\item Decisions must be ratified afterwards by means of small scale social choice (normally involving non-anonymous voting), performed by bodies of agents who have been put in place for limited periods of time by social choice mechanisms implemented by way of anonymous voting. 
\item The primary responsibility for the (consequences of effectuating the outcome of the) decision remain with the agent who took the decision, who is by default assumed to expect ratification.
\end{itemize}
\item Confidential boardroom voting with individual voting outcomes open to all participants may be in some cases be considered an instance of decision taking, whereas the voting that takes place with national elections in so-called democracies does not. Indeed I will assume that a small group of persons constituting a board, which is engaged in open and personal discussions, can have intentions in spite of possible differences of opinion, whereas a large group of individuals cannot be attributed an intention on the basis of the interpretation of an AVO that came out of a group voting process in which these individuals had been invited to participate.%
\footnote{The social choice processes supposed to be supported by the forms of reasoning as suggested in \cite{Bergstra2011a} may also count as decision taking in my opinion.}
\end{enumerate}

One may contemplate voting as a service (VaaS). In some cases it is admissible that an individual outtasks a voting assignment to another agent. It is implausible, however, that an agent specializes in voting on behalf of other agents on a systematic scale. For this reason I consider VaaS to be an implausible notion. Precisely this observation necessitates making a distinction between voting and deciding in preparation of contemplating DTaaS.

\subsection{Promise issuing}
\label{Promises}
Promising, that is,  issuing a promise, is an action comparable to but yet different from taking a decision.  This matter needs to be analyzed at this point because it impacts on the plausibility of DTaaS. For the notion of a promise I refer to \cite{Burgess2005,Burgess2007,BergstraBurgess2008,BergstraBurgess2014}. As discussed in \cite{Bergstra2011d} there is a connection between promise and decision. This connection would be simpler and more symmetric  if a promise were understood as an act of promising. Because that convention would deviate too much from common usage I will not primarily understand a promise as an act of promising but rather as its outcome. A promise is the tangible or intangible outcome of an act of promising and promise issuing (PI)  is performing the act of promising. 

There is remarkable complementarity between promise issuing and voting. Both voting and promise issuing produce an outcome from which further consequences may result. Voting involves choice, but it need not be the expression of an intention or of an expectation. Promise issuing involves intentions and expectations but it need not involve  choice. 

I will distinguish promise issuing from promise making, with promise issuing referring to the action, and promise making referring to a more comprehensive process of which promise issuing constitutes the concluding phase. Promise making may include preparatory steps, mainly for designing the form and the content of the promise
(also called promise body in \cite{BergstraBurgess2008}). 

Promise issuing is close to decision taking but it differs in important ways. I suggest the following relation: every decision outcome is a promise but not the other way around.%
\footnote{The viewpoint that each decision is an instance of promise issuing derives from the formal definition of a promise in \cite{BergstraBurgess2008}. For different definitions of promise issuing the relation with decision may work out differently.}
Promise is more general than decision outcome because: (i) the promiser (comparable to the decision taker) need not have any expectations concerning the consequences of effectuation of the promise, (ii) the promise may be intangible (like a mathematical entity, which stands for a shared cognition) while a decision outcome must be tangible, (iii) the role of a promiser is immaterial, (iv) if a promise has an intangible form, for instance two or more agents remembering that something was said, then the consequences of the promise are mediated via the different components of the promise (that is the cognitive residues of the act of promising as they exist in different agents that were in scope of the promise, including the promiser) and not via the single ``promise'' (viewed as an outcome of promise taking), (v) the scope of a decision outcome is merely its  readership as prescribed by the decision taking protocol at hand, whereas the scope of a promise (see \cite{BergstraBurgess2008}) is more immediately determined by the act of promising involved.

The reason to consider promise issuing in some detail arises from the following observation:
``promise issuing as a service (PIaaS)'' is a problematic (that is, not necessarily plausible) concept because (i) it is unclear how promise issuing can be delegated to another agent, (ii)  it is unclear what it might mean to be more capable of promising on behalf of a promiser than the promiser is him/herself.%
\footnote{``Marketing as a service (MaaS)'' is a plausible notion, and some forms of  MaaS might  be viewed as an instance of PIaaS. The difficulty with understanding PIaaS is immediately connected to the difficulty of explaining the precise role of promises. The abundance of promises ``in the real world'' seems not to be based on the existence of a definite semantics of the term promise but rather on its absence.}
I conclude that although DT is a subcategory of PI, for PI in general PIaaS is a problematic notion. At the same time I intend to maintain that DTaaS is a coherent (potentially plausible) concept. 

So it must be the case that especially for cases of PI that do not qualify as DT provision as a service is implausible. A criterion that separates promises for which PIaaS is implausible from decisions is the concurrent creation of obligations for the promiser.  Some promises create obligations for the promiser.%
\footnote{According to \cite{Burgess2005,Burgess2007,BergstraBurgess2008} promise issuing need not involve the creation of an obligation for the promiser. In particular promise issuing by automated agents must be viewed as autonomous action.} Obligation creating promise issuing is not an instance of decision taking. It appears that ``obligation creating promise issuing as a service (OCPIaaS)'' is an implausible notion. That implausibility renders PIaaS problematic, but not DTaaS.

\section{DT as an (organizational)  feature (DTaaF)}
\label{DTaaF}
Besides interpreting decision as an act rather than as an outcome I will consider it to be a feature of organizational structure, amenable for analysis in terms organizational design and architecture,  rather than as an aspect of human behavior primarily amenable to empirical investigation. DTaaF is a perspective on DT which allows to assess from an external perspective which activities count as DT and which do not. DTaaF also allows the view that an organization may be transformed as to feature more (or less)  DT, that is to have more (or less) frequent occurrences of activity classified as DT.

Decision taking is a concept that emerges from attempts to modularize the behavior of existing or imagined organizations, rather than a concept which emerges when considering the behavior of single individuals or the collective behavior of groups of individuals.%
\footnote{In psychological research it seems to be taken for granted that many forms of behavior, including most forms of human choice, may be considered  instances of decision making, which is usually not distinguished from decision taking. 
I disagree with that usage of the language to the extent that in my view when making a choice is observed no decision taking is necessarily  implied. Choice takes place whenever an agent acts in a setting where the agent was aware of alternative actions that it might have performed instead in full compliance with operational constraints of the setting. An animal may also perform choices even if it is harder to determine the meaning of awareness in that case. That an animal is able to take decisions is implausible according to \cite{Bergstra2011d}. }
I will follow \cite{Bergstra2011d} with the following assumptions: 
\begin{enumerate}
\item A decision is an act of decision taking with actor (agent), time and place as required coordinates.
\item A decision must produce a decision outcome which is a tangible item for instance a text.
\item The effectuation of a decision outcome leads to the consequences of that decision outcome. These 
consequences may be referred to as  the decision effect. Whereas the decision outcome is in existence immediately after and as a direct consequence of the decision, the decision effect may be unpredictable, its
identification may be controversial and its coming about may take much time.
\item Every decision is taken by an agent with the intention that effectuation of the resulting decision outcome has the expected consequences; it depends on the role of the decision taker who needs to take care of effectuating the decision outcome.
\item If agent $a$ takes a decision that is action is performed in the context of $a$ playing some specific role, without that role being known or specified $a$'s production of a decision outcome amounts to no more than a mere speech act, 
\item Rather than the choice between different options, the production of an intermediate product (the decision outcome) is the primary feature of a decision.%
\footnote{Typically a car driver makes choices which do not qualify as decisions when handling the controls of the car while in motion. However, the step to install a car kit for a mobile phone, or the step to buy a new map or a (new) navigation device, may or may not be the consequence of a decision. For instance if a form is created and signed to instruct a car dealer to install a new car kit, that form may be understood as a decision outcome of the car owner's decision to acquire a new car device. But if besides buying food, fuel, and newspaper, a map is bought because of its fresh look and feel, the latter action may not count as having been done as an effectuation of a decision outcome.} 
\item In the absence of a decision outcome, the effectuation of which brings about the intended consequences (as expected by the decision taker), making a choice between different behavioral options amounts to a choice and a choice may be made in the absence of a decision.
\item The most obvious alternative to a decision is not to take that decision (not to take another decision).
\item Decision taking is a concluding subprocess of decision making, decision making primarily involves 
the preparation of decision outcomes. Decision making is embedded in a larger decision making process which also involves protocol actions that do not influence decision outcomes.
\item A decision terminates an episode of indecision.  
\item A decision is an action but it need not be a decisive action, while a decisive action need not be or involve a decision. A decision is a decisive action only if in hindsight it appears that effectuation of the decision outcome had decisive consequences.
\item Making up one's mind in preparation of an action is not an instance of decision taking.
\item A command (also called an imposition in~\cite{BergstraBurgess2014}) is the outcome of an act of commanding  (command issuing) and it is like a decision but it indicates in addition which agent has to act. 
If the command is tangible it comprises a decision, but if it is intangible the command may be a speech act or a gesture in which case it will usually fail to count as a decision. Thus: command issuing that leads to a tangible command (outcome oriented command issuing) produces a decision at the same time, whereas on the fly command issuing produces an intangible command only which does not count as a decision.
\end{enumerate}
In addition to what was stated in \cite{Bergstra2011d} it is required that a decision outcome is largely self-explanatory. A mere bit (0 or 1, usually termed ``yes'' or ``no'') is insufficient in the absence of an unambiguous reference to a rendering of the question to which that bit is supposed to constitute an answer.%
\footnote{When voting an agent may choose ``yes'' (or ``no'') unaware of its implications or meaning. In our understanding of the act of voting a vote may be cast by an agent without having any intentions in mind.}

\subsection{Remote method invocation as an analogue}
My perspective on decision taking is derived from a perspective on activity found in the area of computer programming.  When observing a machine effectuating an instruction sequence one sees no more than a sequence of steps. That the instruction sequence has been written in a so-called structured notation is a matter of conceptual organization at a higher level of abstraction. 

For instance it might be stated that a certain sequence of steps is best generated by effectuating a method call from an appropriate class thus providing the additional judgement that methods and classes are useful means of structuring in the case at hand. Seen from the perspective of an effectuating machine it may be preferable that a particular method is performed by another machine, which comes close to it being outtasked to that (other) machine, thus leading to so-called remote method invocation, or in terms of our topic: method effectuation as a service (MEaaS).

Rather than viewing decisions as very particular actions taken by agents or by organizations, decision may be seen as a way to structure organizational behavior. For instance it may be ``decided'' (at the level of organizational design) that a particular class of actions must always to be preceded by decisions so that the effectuation of an action from that class  can be understood as the effectuation of a decision outcome to that end. That is useful only if a protocol for decision taking, as well as for the preparatory part of decision making is laid down as well.

From this perspective it is plausible that if an organization $U$ plans to make use of DTaaS it may need  a preparatory phase for structuring its operations in such a way that DT takes a more prominent place to begin with. In particular DT may then take place as a thread amongst other threads in a multi-thread run by way of strategic interleaving by some managing body.%
\footnote{Changing an organization into a mode of operation where decision taking is more prominent has virtues outside the context of DTaaS. Achieving such transformations may be valuable if increased accountability and transparency are sought. Of course the converse holds as well: turning decisions into mere choices may be helpful if accountability, traceability, or transparency are to be diminished.} 
The transition towards DTaaS where some external agent $s$ provides a similar thread as a service to $U$'s management body may be understood as temporary thread migration. 

DT as provided by an internal service must be monitored internally from the perspective of DT demand management. Only if it is known what non-DT threads expect from a DT thread it can be agreed what to expect from an external provider of a DT thread.

\subsection{Decision-low operation as an organizational feature}
By outtasking DT to other organizations (and using DTaaS) 
 an organization may become ``decision-low'', that its its operation
is rather independent of its own internal DT activities. Considering organization X its low-decision feature may come about in various ways:
\begin{itemize}
\item not yet having developed a tradition of internal DT,
\item abandoning roles from which DT is plausible,
\item abandoning business processes and tasks that require DT,
\item outtasking DT,
\item adapting processes tasks in such a way that choice and action determination replaces decision taking.
\end{itemize}
Below in~\ref{prfuDTaaS} I will outline possible advantages of decision low-ness for an organization.

\section{On services and service casting}
\label{Osasc}
A service is defined as an identifiable and intangible activity that is the main object of a transaction in \cite{SommersBSEW1995} (see also \cite{ChungMcLarney2000}). I will use a somewhat more detailed definition
of a service, useful in particular to separate services from products, that was used in 
\cite{NiessinkVliet2000}.
In \cite{NiessinkVliet2000} the distinction between products and services is clarified by means of an extensive quote from \cite{ZeithamlBitner1966}. In brief a service is sold by one party to another, but as opposed to product, it is intangible (as an entity), heterogeneous (as a concept), it is perishable (non-enduring), and it is necessarily consumed and produced concurrently. According to  \cite{NiessinkVliet2000}, however, services may carry products along and products may carry services along, the difference being a matter of gradation rather than a very sharp one. One may perhaps simplify these requirements to the condition that a service is a process which can be provided against compensation by a third party.

\subsection{Casting activities and products as a service, a catalogue of examples}
\label{SCexamples}
In \cite{DoerrBVH2010} ``music as a service (MaaS)'' is explained to be the transfer of music in digital form, by way of streaming without transferral of ownership of the data. MaaS is seen as an instance of content as a service (CaaS). Now if one listens to ordinary live music the suggestion that ownership is transferred 
has never been present, and that suggests that live music has always been music as a service. However, in order to grasp the phrase ``music as a service'' one needs to understand that live music is not meant with that phrase as it would be odd to look for a new name for such a classical phenomenon. For that reason the phrase MaaS lives in another world, most plausibly the digital world, where music as a a product is known, and music as a service may be a novel phenomenon. More can be said, however: the provider of live music cannot concurrently serve different customers with different tunes, whereas a digital service provider is supposed to be able to serve different clients at the same time in ways more flexible than mere broadcasting. This concurrency criterion on service provision is absent in the discussion of \cite{NiessinkVliet2000} but it seems to be a helpful addition to understanding what is amenable to service provision and what is not.

In \cite{PauwelsACLRSWW2009} business dashboards as a service are discussed. As far as I can see the authors fail to explain the plausibility of service casting in this case. Instead they merely analyze the virtues of dashboards (implemented as an interactive tool for business data visualization) for management purposes as well as the need for further research about them. The phrase ``dashboards as a service'' suggests the presence of a third party collecting a client's business information and extracting from that a clever abstraction which is provided to the client in the form of a dashboard, which itself is an internet service. This is a plausible explanation of what ``dashboards as a service'' might amount to, irrespective of the extent to which that constitutes a realistic IT market opportunity at this moment in time. In the discussion of \cite{NiessinkVliet2000} from which we have already quoted criteria for the distinction between product and service, the presence of a third party arises from their condition that a service can be sold by one party to another.

In \cite{LangSchreiner2011} the usage of ``(security) policy as a service'' is explained in the context of cloud security. This is a convincing example of service casting. A security policy is intangible and non-product like but its real time provision by a third party (and hosted in the cloud) may be understood as a novel feature.

In \cite{RosenbergLMCD2009} one finds an elaborate description of (software) ``composition as a service'' (abbreviated as CaaS in \cite{RosenbergLMCD2009} but here as SCaaS in order to avoid confusion with  ``content as a service''). SCaaS is presented as a phenomenon in the context of software as a service (SaaS). SCaaS differs from SC by its provision by a third party which  justifies service casting in this case. Software composition satisfies all requirements in the entities traded as services as formulated in \cite{NiessinkVliet2000}.

In \cite{BhardwajJJ2010} a description of cloud computing is given with a focus on ``infrastructure as a service'' (IaaS). IaaS is considered the most generic service provided by a cloud in addition to PaaS (platform as a service, see also
\cite{Lawton2008}) which is committed to some fixed brand of systems software, and SaaS (software as a service) which in addition implies a commitment to an application area and a family of software products applicable to that area from which services may be composed automatically on demand. The phrase IaaS is difficult to grasp because infrastructure primarily refers to a product rather than to a process. It seems to be more precise to speak of ``infrastructure behavior as a service''  (IBaaS) instead. 

In \cite{AalstAHPS2009} flexibility as a service (FaaS) is proposed. Although flexibility is no more than an attribute of a service offering, this instance of service casting is based upon the fact that principles of service composition can be provided in such a way that flexibility results. Remarkably, ``services as a service'' (SVaaS) becomes meaningful if it implies a focus on service construction by means of systematic composition in real time at the provider side. As \cite{AalstAHPS2009} points out,  FaaS is an objective which can be obtained through what I just termed the SVaaS perspective.

Software as service is currently the most well-known instance of service casting, by which all other instances seem to have been inspired.%
\footnote{If software is identified with ``families of polyadic instruction sequences'', an identification which I consider to be adequate, it becomes reasonable to understand software as a mathematical entity, which is intangible rather than tangible. Of course the physical representations of mathematical objects are tangible, but the mathematical objects themselves are not, or not necessarily. Thus I will identify computer software with ``(conventional) physical representations of families of polyadic instruction sequences'', thereby rendering software tangible.} 
I will now discuss this example of service casting in some detail.

Finally an interesting case is ``politics as a service (POLaaS)''.%
\footnote{The late Dutch politician Pim Fortuyn made headlines in the Netherlands with his slogan ``at your service'', 
which may be understood as his intention to provide POLaaS to the Dutch public. The extent to which POLaaS is a 
credible option that escapes from the problems of plain power oriented populism still 
puzzles Dutch political commentators 
ten years after Fortuyn's violent death in 2002.}
POLaaS (often written PaaS, which unfortunately collides with ``platform as a service") has a formidable presence on the internet.

\subsubsection{SaaS, SEffaaS, SEaaS, STaaS, and SPaaS}
In spite of software constituting a product (tangible good) rather than a process (intangible good), SaaS (software as a service, see \cite{TurnerBB2003}) is at present the most prominent instance of service casting.

Because software is effectuated when used, ``software effectuation as a service'' (SEffaaS) 
makes perfect sense, (with AHaaS for ``application hosting as a service'' as the more common name).  SaaS goes beyond SEffaaS however, as it also comprises the on demand and real time composition of services (as provided by software being effectuated), and it may include the provision of software from a remote server. SEffaaS and SEaaS ([Software Engineering]aaS) are both included in SPaaS (software process as a service), which goes beyond SEffaaS and SEaaS (by taking all phases of the software life-cycle into account. SPaaS also comprises the service area STaaS (software testing as a service, also known as TaaS) and SDaaS (software development as  as service). Another option is PRSaaS for providing software as a service.
T
\subsubsection{(Provision of software) as a service =\\ provision of (software as a service)?}
The background of all service casting expressions XaaS might, if only as a thought experiment,  be understood as follows: (i) XaaS is a service, (ii) every service H is identified with providing H, (which is abbreviated to PR H). Thus: XaaS = PR (XaaS), 
(iii) associativity of bracketing is used PR (XaaS) = (PR X)aaS. In the case of software:

Software as a service = providing (software as a service) = (providing software) as a service.

Identification of a service casting with its provision may be considered imprecise, and that leads to contemplating inverse-provision as a kind of inverse operator for service casting. If inverse-provision denotes the  abstraction from the phenomenon of the provision of a service to the abstraction consisting of its underlying service then one obtains the equation: 

XaaS = inverse-provision ((provision of X) as a service).

This explanation of service casting would imply that it renders every service casting plausible, which is itself an implausible state of affairs, and for that reason I will not take it seriously.

\subsection{Plausibility of service casting}
I will use concept as a category that includes, entity, service, process, product, and method. For each concept X, the service casting of X, [X]aaS can be contemplated. Service castings range from implausible to plausible. X satisfies the so-called service casting preconditions of Paragraph \ref{pcscp} below if and only if [X]aaS is a plausible service casting.

The question arises which cases of service casting are plausible. That question is seemingly paradoxical in the following sense: if X is a service then XaaS makes little sense because [X]aaS suggests that $X$ is seen from the service perspective, instead of its ``natural'' perspective. 

\subsubsection{Implausible cases of service casting}
\label{icsc}

If X is not nearly a service then it may  defeat being viewed from a service perspective. Here are some examples. None of the following service castings seems to make sense: ``jogging as a service'', ``meaning as a service'',  ``gold ownership as a service'', ``medication as a service'', ``bicycles as a service'', ``hardware as a service'', ``sleeping as a service'', ``breathing as a service'', ``flying as a service'',   ``consciousness as a service'',  ``being born as a service'', ``self-confidence as a service'', and ``dining as a service''. 

\subsubsection{Unclear cases of service casting}
There are unclear cases as well. At this moment I have no judgement available on the plausibility of the following
potential service castings:  ``choice as a service'', ``social choice as a  service'', ``owning as a service'', ``spying as a service'', ``swimming as a service'', ``natural numbers as a service'', ``walking as a service'', or ``happiness as a service''.%
\footnote{In contrast ``well-being as a service'' (often referred to as wellness service) is plausible.}

\subsubsection{Service casting obstacles}
Service casting obstacles are preconditions on a concept X that render XaaS implausible under all circumstances. Two such conditions can be identified at this stage: X is a service already (a violation of this condition will be labeled as weak implausibility of the service casting at hand), and non-disposablity, that is after outtasking or outsourcing X from $U$, an unacceptable loss of identity of $U$ has occurred.

\subsection{Service casting preconditions}
\label{pcscp}
The following conditions on X must be met for [X]aaS to be plausible:%
\footnote{Plausibility is a semantic notion purely related to the amenability of X for being offered as a service, or for suggesting a clear interpretation of [X]aaS which  is to be distinguished from usefulness of [X]aaS, once it has been accepted as a plausible concept, for its consumer and producer.} 

\begin{description}
\item{\em Ontological constraints.} Preferably X   satisfies the criteria listed in 
\cite{NiessinkVliet2000} for services. To meet those criteria X needs to be intangible, and non-enduring,
 and P's production and consumption must be necessarily concurrent. Preferably X is a kind of activity rather than a kind of entity. If X is tangible the following clause applies.
\item{\em Service casting preconditions for product classes.}
If X is  product-like (tangible) rather than service-like (and therefore intangible) 
and if [-]aaS is instantiated to XaaS, then the following conditions must be met:

\begin{enumerate}
\item XaaS will provide processes P that usually represent the usage of X (or of instances of X),%
\footnote{If XB represents the behavior of an X (or of an instance of X) then XaaS is understood as XBaaS by default.}
and,
\item if a process P representing the usage of X (or of an X) can be provided through the internet, services as meant by XaaS will be delivered as a webservice, and several customers may be provided the same service at the same time.

\item if P is  most likely is to be provided via the internet, and if normal effectuation of the process X does not involve human operation then XaaS will definitely not involve human operation.%
\footnote{In the case of SaaS (software as a service)  this constraint explains why SaaS will not extend to ``software engineering as a service'' (SENaaS), a ``product''  involving human engineers that has been delivered by the software industry for many years.}
\end{enumerate}
\item{\em Significant deviation from normal casting.}
The concept X is not normally seen as a service.%
\footnote{I propose to call ``online banking as a service'' weakly implausible because online banking is normally considered a service.}
A violation of this precondition amounts to weak implausibility. 
Weak implausibility does not imply incoherence, it merely implies service casting is applied in an implausible case. 
From these considerations it follows that casting activity, or entity, 
X as a service requires that the audience for the phrase ``X as a service (abbreviated XaaS)'' 
accepts the following implicit ambiguity: 
\begin{itemize}
\item  X must not a service by default.  
But the difference between the default role of X and its role after being cast as a service should not be too large either. 
If it is too large one arrives at implausible service castings, some of which have just been listed. 
\item The more distant XaaS is from the normal role of X the more successful the notion XaaS may become. 
Probably SaaS has acquired significant popularity just because software is commonly 
understood as a product (and not as a service).
\end{itemize}
\item{\em self X is the normal (default) case.}
If X is normally a service it can be placed in the context self[-] thus obtaining ``self[X]''.%
\footnote{For instance self hairdressing.} If it is explicitly meant that X is understood in its default meaning that can be expressed with ``X as usual''.%
\footnote{For instance: ``self hairdressing'' is probably cheaper than ``hairdressing as usual'', thus avoiding ``hairdressing as a service'' which is weakly implausible.}
\item{\em Demand management phase is an option.} As a preparation for a XaaS phase for it prospective consumer $U$  may need to structure its own X processes in such a way that X is performed as self[X] (by $U$), though in a separate unit which is monitored and managed with principles of demand management. Once this managed self[X] process is in place
outsourcing or outtasking X to an appropriate service provider is easier to achieve.

\item{\em Disposability.}
It must be the case that stripping X from a unit (that is terminating self[X]) need not 
necessarily imply a loss of its identity.

\item{\em Concurrency potential.}
Implicit in agenta $a$'s offering a service X to $U$ is that $a$ can, at least  in principle, offer the service X concurrently to a plurality of clients. This form of concurrency can be described by strategic interleaving for multi-threading as outlined in 
\cite{BergstraMiddelburg2007}. That concurrent offering of X is an option is a property of X and it can be understood as a service casting precondition.

\item {\em Potential for co-creation of the service by provider and consumer.} It is sometimes proposed as a key aspect of services that these are co-created by producer and consumer. 
\end{description}

\section{DTaaS}
\label{DTaaS}
After considerable preparations a ``formal'' (in the sense of ``official'') introduction of DTaaS is now enabled. The introduction splits in four parts: (i) creating DTaaS as a (meaningful) concept, (ii) a survey of DT specific aspects of DTaaS instances, (iii) a survey of potential advantages of DTaaS for its consumer and provider, and (iv) a survey of  risks and potential disadvantages for DTaaS.
\subsection{DTaaS: a definition}
As a concept, DTaaS is the result of service casting applied to the concept DT. 
For this result to be well-defined compliance with the 
various service casting preconditions needs to be checked. This matter will be addressed  in Paragraph \ref{DTmeets} below. 

Each instance of DTaaS involves a unit and an agent. DTaaS refers to a particular kind of sourcement: DTaaS occurs if an agent $a$ provides DT as a service to a unit $U.$ In the case that $a$ provides DTaaS to $U$, the decisions taken by $a$ are taken on behalf of $U$ and have the same status as decisions taken by $U$'s management. 

Much can be said about instances of DTaaS which is specific for DT as a concept. For instance that
this sourcement is likely to have arisen by outtasking 
a part of $U$'s DT to $a$, where outtasking refers to a temporary transfer of an activity (task) to another unit (the unit containing $a$) without a corresponding transfer of sources. In Paragraph \ref{DTaaSmas} a range of further observations concerning instances of service casting DT, which are specific to DT rather than to service casting in general, is made.

\subsubsection{DT meets service casting preconditions}
\label{DTmeets}
DT is intangible, and it is not normally cast as a service. Further self DT is the default situation and demand management for DT is an option, though in order to understand that option one probably needs an awareness of the notion of DTaaS
already. DT is disposable in some cases and a DTaaS  provider may concurrently provide DT services for different clients. By developing adequate DT protocols, a DP task which may be shared with the client, co-creation of DT services between provider and consumer may be achieved.

\subsubsection{DTaaS avoids service casting obstacles}
DT is not usually considered a service, and outtasking DT, in whole or in part, need not necessarily degrade a unit's identity. In \ref{DFO} a listing has been made of the activities that differ from DT and that will remain after outtasking DT.

\subsection{DTaaS: DT specific aspects}
\label{DTaaSmas}
The purpose of this section is to specify the result of service casting of DT,  and more specifically of DT$_I$ (leading to DT$_I$aaS) for some decision interface $I$, in greater detail than merely stating that it results from service casting.

\subsubsection{Which decisions can be delegated?}
Which decisions can be delegated to a service provider? Because it is hardly conceivable to hand over ``all DT''  to an external agent, there is no way around some form of modularization. To that end one may take advantage of the concept of a decision interface (DI), which collects a coherent family of issues or topics about which decisions may be taken. Elements of a decision interface are sometimes called decision rights, but that is asymmetric as these elements may equally well be understood as decision obligations (or more neutral, decision tasks, decision options, or decision patterns). The notion of a decision interface is symmetric. It is understood that an agent having regular DT as its task may collect all of its potential decisions in an overall DI which is modularized (decomposed) as a combination of disjoint sub DI's. 

Given a decision interface $I$, DT$_I$ represents DT concerning decisions in $I$, DM$_I$ stands for DM concerning decisions in $I$ and DP$_I$ concerns DP limited to decisions in $I$. We have the following $I$-specialized equation 
``DM$_I$ = DP$_I$ + DT$_I$.''

 In \cite{Bergstra2011d} it has been argued that DT requires that the
	the DT agent operates in a clearly determined role. Thus some role must be provided, at the least this role
	is the following ``external DT agent on behalf of $U$'', or  ``external DTaaS provider for $U$'', 
	or simply ``external decision taker for $U$''. The role name is likely to be linked to the decision interface 
	for which the role has been sought.
	
	 It is an existing practice that external DTaaS is sought for brief episodes and dedicated to a single issue or 
	to a very confined range of issues,  the role of an external DTaaS provider has more intrinsic names, 
	for instance a jury, if a prize 
	is to be awarded on behalf of $U$,
	
	 it is worth mentioning some further examples of  role names for thematic 
	 episode driven instances of DTaaS (abbreviated as EdDTaaS) 
	with limited scope: (i) a 
	mediator if mediation is sought for a conflict between subunits within $U$ (and 
	DTaaS is applied because $m_U$ may not be considered sufficiently impartial, or may not have the required 
	authority or trust base within $U$), (ii) a court if $U$ has sought the intervention of a legal process, (iii) 
	some specialized authority (e.g. the EFSA, European Food Safety Authority) which is asked to intervene, (iv)
	an active private banker who may trade on behalf of its clients, (v) a mountaineering guide is in 
	charge of a well understood part of his/her client's DT during a trip, (vi) an auctioneer decides about 
	transactions between parties who have each agreed to stay within the protocol prescribed by the auctioneer.

\subsubsection{What sourcing options exist for DT?}
 Having stated that an instance of DTaaS is a sourcement of a particular kind, it is useful to consider in some detail which sourcements may allow an external agent $a$ to perform DT (or DT$_I$) on behalf of the management $m_U$ of some unit $U$? Here are some observations concerning that matter.
	\begin{itemize}
	\item Unit management may be self-sourcing for DT, that is $m_U$ performs DT (in other words it applies self DT). This is most common and I will
	assume that it is the case by default.
	\item DT may have been delegated by  $m_U$ to one or more subunits at a lower 
	level in $U$'s management hierarchy. 	In this case $U$ is self-sourcing for DT but $m_U$ is not.
	\item $U$ may sometimes distinguish between governance ($g_U$) and management ($m_U$), in 
	which case it is the role of $g_U$ to delegate DT to management functions within $U$.
	\item if $g_U$ temporarily delegates DP to an interim manager $im_U$ instead of to $m_U$, $U$ is still 
	self-sourcing for DT because $im_U$ operates under the responsibility of $g_U$.
	\item $U$ may be temporarily non self-sourcing for DT. 
	That means that DT has been temporarily delegated (outtasked) by $g_U$ to an 
	agent $a$ operating outside $U$. In this case $a$ acts as a service provider for $m_U.$ The authority and 
	steady support of $g_U$ is
	required to assure that $a$ is in a decision taking role rather than merely in a consulting role 
	where it may at best suggest potential decision outcomes to $m_U$. 
	
	As long as $g_U$ keeps the
	outtasking relationship between $U$ and $a$ in place $m_U$ accepts the decision outcomes of $a$
	decisions as if these were decision outcomes brought about by its own decision taking. It is essential 
	to specify in advance which types of decisions constitute part of the DTaaS agreement, 
	because $m_U$ cannot undo $a$'s decisions any easier than its own decisions.
	\end{itemize}
	
\subsubsection{Which sourcement transformations exist for DT?} 
In \cite{BDV2011c} the notion of a sourcement transformation has been examined in detail, with a particular focus on the following transformations: outsourcing, insourcing, backsourcing, and follow-up outsourcing. Without further analysis I will postulate that corresponding transformations can be found if only tasks but no sources are being  transferred: outtasking, intasking, backtasking and follow-up outtasking. For the case of DTaaS the outtasking transformation is of particular importance.
	\begin{itemize}
	\item DT cannot be outsourced. This is so because outsourcing of DT (from a state where it is performed by $m_U$)
	 involves moving $m_U$ (or a part of it) outside of $U$, which cannot be done unless
	the characteristic feature of its being the management of $U$ is lost. An artificial option which I will discard 
	as being unconvincing is to give members of $m_U$ part-time positions in an 
	insourcing unit $U^{\prime}$ outside $U$.
	
	Another artificial opting that is mentioned only is that given the decision interface $I$, the body $m_U$ is 
	decomposed as $m_U = m_U^{d,I} \cup m_U^{nd,I}$  into 
	disjoint groups $m_U^{d,I}$ (involved in decision taking about decision patterns in $I$)
	 and $m_U^{nd,I}$ (not involved in decision taking concerning decision patterns in $I$), and to assume that 
	 $m_U^{nd,I}$ is outsourced by $U$ to $a$.  This option is artificial because it casts doubts on the status of 
	 $m_U$ as a managing body in the original situation.
	\item outtasking of DT is possible, a consequence that being that the tasks of $m_U$ are temporarily reduced. 
	\item Task externalization is not an option for DT. Indeed it is against the required autonomy 
	of $U$ as being an independent unit if DT is transferred to an external agent on an indefinite basis. Thus the
	feature of outtasking that it comprises a temporary transfer of tasks is essential in the case of transferral of DT.
	\item If DT has been outtasked to agent $a$, that agent is called an external decision taker on behalf of $U$.
	\item Because DT cannot be outsourced and DT is a part of DM, only partial outsourcing of DM is possible.
	In particular only partial or full  outsourcing  of PM is possible. 
	\item Typically DP or parts of it can be outtasked to a so-called consultant. 
	\item By definition DT cannot be outtasked to a consultant (in its role as a consultant).
	\end{itemize}
	
\subsection{Possible reasons for engaging in DTaaS}
\label{prfuDTaaS}
DTaaS is a (potentially) meaningful concept in the sense that it may open a novel perspective on organizational architecture. DTaaS seems to be uncommon except in cases where quite dedicated DT processes are outtasked and often in a case by case fashion only. 

DTaaS will be initiated by the outtasking unit $U$, or rather by its governing body (subunit) $g_U$ 
which will make use of the DT services of agent $a$ which may be a part of unit $U^{\prime}$ instead 
of having that part of DT self-sourced for $U$ by $m_U$. For enacting this outtasking transformation  
$g_U$ needs plausible reasons. 
These reasons will differ from case to case but some general remarks concerning these reasons can be collected:
\begin{itemize}
\item DT provider $a$ may be less vulnerable to the effects of stress created by an involvement in DT concerning $U$ than $m_U$.
Even $a$'s authority or trust base may exceed that of $m_U$.

\item DT provider $a$'s awareness of the consequences of possible decisions to parties outside $U$ may 
exceed that of $U$. 

\item DT provider $a$ may include highly skilled DT specialists who are especially competent in: (i) timing of DT, (ii) prediction of the effects of implementing potential decision outcomes, either tactically or strategically, or both, (iii) managing the core of DP preceding DT, (iv) evaluating the quality of preparatory DP activities may exceed that of $m_U$,  (v) drafting decision outcomes, (vi) communicating decision outcomes.

\item DT provider $a$ may be able to apply process models to DT which it has used elsewhere and by doing so perform at a higher level of competence compared to the competence level provided by the DT function within $U$. For instance $a$ may be well placed to deal with (that is to guarantee) abstraction (preventing information from leaking to third parties) and encapsulation (preventing external interference)  for the decision process.%
\footnote{Abstraction and encapsulation are notions from process theory (\cite{BaetenBastenReniers2009}) which find an application in many decision making process.}

\item Every decision may be understood as the effectuation of a decision taking thread, itself the result of putting into effect an underlying instruction sequence (e.g. see \cite{BM12a}). Effectuating such instruction sequences%
\footnote{In particular these instruction sequences may involve conditions written in the notation of short-circuit logic, which is semantically analyzed by proposition algebra (see \cite{BergstraPonse2010}).}
may be a particular competence of $a$.

\item As argued in \cite{Bergstra2011d} DT may involve multi-threading (see \cite{BergstraMiddelburg2007}) in the case of DT for $U$. As $a$ may be providing
DTaaS services for several different clients simultaneously $a$ may be dealing with hierarchical multi-threading (see
\cite{BergstraMiddelburg2006}). More efficient methods for strategic interleaving of hierarchical decision threads may be available to $a$ than to $m_U.$ 

\item By outtasking DT $m_U$ arrives at a situation where it can concentrate on putting decision outcomes into effect. 
In that case $m_U$ may operate in an almost decision free fashion. 
Once decision outcome effectuation has become standard practice $m_U$ may terminate outtasking DT or part of it.

\item For $U$ it may be helpful to forget about short term tactical DT in order to fully concentrate on long term strategic matters. The opposite may hold as well: then $m_U$ makes use of DTaaS provided by $a$ hoping that $a$ is able to deal with long term and strategic matters while $m_U$ is capable of operating with full concentration on tactical and sort term issues of vital importance to $u$'s existence.

\item DTaaS can be used to reduce the burocracy in the outtasking organization. Burocratic simplification can conceivably be achieved by training staff to perform a significant amount of action without the need for DT processes, thus reducing the volume of DT and outtasking a part of the remaining DT may further reduce the dependency on DT. Once DT is out of the way business process optimization and automaton can both be used to streamline an organization's workflow.

\item DT provider $a$ shares the responsibility of (the effectuation of the decision outcomes of) 
its decisions with the unit $U$ to which
the service is provided. The legitimacy of $a$ having these responsibilities has several sides, but a sufficient criterion is that $g_U$, which is in charge of ultimate responsibilities for $U$, can argue convincingly that it is in $U$'s best interest to outtask DT to $a$. In particular if an emergency of a particular kind has to be dealt with, $a$'s competences to deal with the variety of issues connected with that particular kind of emergency may be valued of higher importance than conformity of $a$'s own intentions with achieving some or all of $U$'s strategic objectives.
\end{itemize}

\subsection{DTaaS risk analysis}
DTaaS is a thought experiment rather than common practice. DTaaS for short episodes or limited to relatively confined decision classes is common practice, but DTaaS for the full width of $m_U$ decision rights seems to be unusual. Nevertheless, even as a thought 
experiment DTaaS is amenable to risk analysis. 
There is a plethora of potential risks that DTaaS carries with it for its consumer. I will mention only a few aspects of the risk analysis.
\begin{itemize}
\item There may be a risk of vendor lock in when $U$ makes use of DTaaS provided by $a$ If DP is outsourced to $a$ there is a risk that $U$ may not be able to realize backsourcing of its DM activities, thus becoming dependent of $a$. This
may be a reason for not outsourcing DP to the same service provider as the one to which DT is being outtasked. It even
suggests that outsourcing DP brings with it a higher risk of vendor lock in than outtasking DT may do.  If only DT has been outtasked to $a$, $U$ must make sure, and communicate that backtasking DT from $a$  is an option at regular instances of time.

\item Like in the case of an external consultant $U$'s appreciation of $a$'s services may degrade. In that case  $a$'s staff may withdraw from its involvement in a DTaaS sourcement  if its authority has become problematic and its activity is less effective.

\item That creates the risk that an occurrence of DTaaS is taken to be a sign that $m_U$ fails to live up to its expectations. In fact $m_U$ seems to carry an obligation to take the majority of its own decisions
itself. DTaaS is about the thought experiment that this obligation is replaced by a less restrictive policy towards the range of task descriptions suitable for a management agent or team $m_U$.

\item Assuming that $a$ provides DTaaS to different units a conflict of interest arises if at the same time $a$ needs to strive for opposite objectives.%
\footnote{Less apparent than an outright conflict of interest is so-called feature interaction (see \cite{KimblerBouma1995}).
If $a$ is involved in providing DT services for different units it needs to analyze feature interaction between its respective decision interfaces. A conflict of intentions is an obvious instance of feature interaction, but more subtle interactions may need attention as well.}
Because $a$ is taking decisions it does so with the awareness of expected consequences (of effectuation of decision outcomes) and with the (delegated) intention to effect these consequences. That intentions can be delegated at all is open for discussion, and I will assume that when acting as a DT provider for $U$, $a$ may pretend intentions (on behalf of $a$) that are not quite its own, provided that these intentions comply with $a$'s best and own intention to maintain a bundle of pretended intentions on behalf of $U$ that best serves $U$'s interests, and to do so in a consistent fashion.

\end{itemize}

\section{Concluding remarks}
\label{Cremarks}
This paper contains a theory of DTaaS and following \cite{BDV2011b} it is fitting to ask about the intended audience of that theory and to ask what  members of that intended audience might get out of it. In the terminology of \cite{BDV2011b} the latter question is phrased as to what ability or which conjectural ability is supposed to result from an acquaintance with this theory. These conjectural abilities are twofold: the ability to acquire a DT task and to start acting as a DTaaS provider (intasking DT, the transition complementary to outtasking DT), and the ability to outtask part of one's DT activities in order to concentrate on other activities which are considered to be of higher importance in some stage. Returning to the matter what constitutes an intended audience for this paper: those potential readers who take an interest in decision taking and have accepted the preparatory analysis made in \cite{Bergstra2011d}.

An analysis of service casting has been provided as well as a demarcation of decision taking separating it from choice, individual voting, social choice, and obligation creating promise issuing. Using these ingredients the conceptual coherence (consistency) of DTaaS has been established and potential advantages of its use have been indicated.

Further work can be descriptive, aiming at finding instances of DTaaS in existence and investigating how that came about and what advantages are actually obtained. Another and perhaps more attractive path is to investigate existing organizations or projects and to find suggestions for first making DTaaF more visible in the organization (or project) and subsequently outtasking a part of DT so that DTaaS is made use of.

As an application of this work DTaaS can be introduced in pilot organizations or projects in order to find out which of the mentioned potential advantages can be achieved in practice, and if so under what circumstances.

\end{document}